\documentclass{PoS}

\usepackage{subcaption}
\usepackage{cite}

\title{Complete off-shell effects for top-antitop + jet production with leptonic decays at the LHC}

\ShortTitle{Complete off-shell effects for  $t\bar{t}j$ production with leptonic decays at the LHC}

\author{\speaker{Giuseppe Bevilacqua} \\
       University of Debrecen and MTA-DE Particle Physics Research Group, H-4002 Debrecen, PO Box 400, Hungary \\
        E-mail: \email{giuseppe.bevilacqua@science.unideb.hu}}

\abstract{A brief summary of the calculation of the NLO QCD corrections to the process $pp \to e^+ \nu_e \mu^- \bar{\nu}_\mu b \bar{b} j + X$ is reported. This provides a complete description of the process of $t\bar{t}+\mbox{jet}$ production with leptonic decays beyond the narrow-width approximation. Off-shell effects for top quarks and $W$ boson decays are fully taken into account, namely all resonant and non-resonant contributions at the order $\mathcal{O}(\alpha_S^4 \alpha^4)$ are included in the calculation. Selected results for total and differential cross sections are shown for the case of the LHC Run I at the energy of 8 TeV.}

\FullConference{XXIV International Workshop on Deep-Inelastic Scattering and Related Subjects\\
	 11-15 April, 2016\\
	 DESY Hamburg, Germany}

\begin{document}

\section{Introduction}

With the start of Run II in 2015, the LHC entered a new stage of operation. Proton collisions are now delivered at the record energy of 13 TeV, compared with the maximum of 8 TeV reached during the previous stage. The first analyses have already started to complement the results of Run I, among which the discovery and characterization of a narrow light Higgs boson was one of the greatest outcomes.
The production of top quark pairs in association with a hard jet ($t\bar{t}j$) is of particular interest in this context. Besides  representing a background for Higgs boson searches in the Vector Boson Fusion and $t\bar{t}H$ channels, it plays an important role in searches of physics beyond the Standard Model (SM). For example, typical signatures of supersymmetric particle decays involve hadronic jets, charged leptons and missing $p_T$, resembling in this way $t\bar{t}+\mbox{jets}$ final states. But the $t\bar{t}j$ process is also an interesting signal on its own. Given that a significant fraction of the inclusive $t\bar{t}$ sample  appears in association with hard jets, the accurate description of this process contributes to a more precise understanding of the dominant mechanism of top quark production at the LHC \cite{Czakon:2013goa,Czakon:2015owf,Czakon:2016dgf}. Last but not least, $t\bar{t}j$ has proven to provide a competitive method for the measurement of the top quark mass based on the analysis of its invariant mass distribution \cite{Alioli:2013mxa,Aad:2015waa}.

\begin{figure}
\begin{center}
  \includegraphics[width=1.0\textwidth]{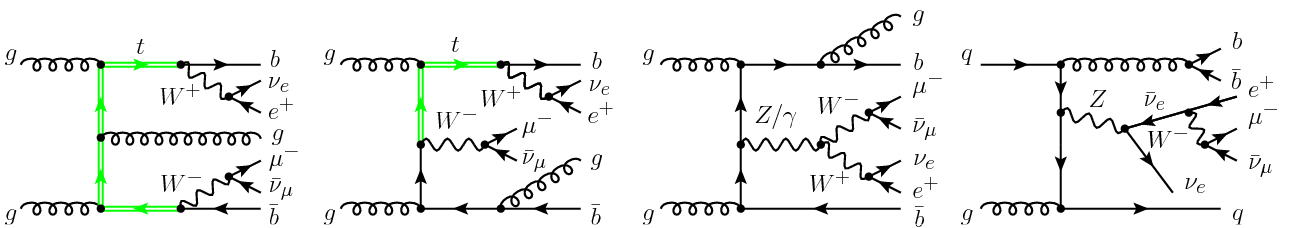}
\put(-382,-12){\small (a)}
\put(-278,-12){\small (b)}
\put(-174,-12){\small (c)}
\put(-70,-12){\small (d)}
\caption{Representative tree-level contributions to $gg \to e^+ \nu_e \mu^- \bar{\nu}_\mu b \bar{b} g$ at the order $\mathcal{O}(\alpha_S^3 \alpha^4)$: \textit{double-resonant} (a), \textit{single-resonant} (b) and \textit{non-resonant} (c,d).  Diagrams (b,c,d) are examples of non-factorizable contributions to $t\bar{t}+\mbox{jet}$ production.}
\label{fig:fig1}
\end{center}
\end{figure}

The lifetime of the top quark is extremely short: any realistic simulation of processes involving top production shall treat tops as intermediate, finite-width states.  Since top quarks decay almost exclusively to a $W$ boson and a $b$ quark in the SM, one of the conceptually simplest final states that provides a complete description of $t\bar{t}j$ hadroproduction is $pp \to e^+ \nu_e \mu^- \bar{\nu}_\mu b \bar{b} j$. By gauge invariance, one must incorporate different kinds of contributions to the amplitude for such final state in addition to the double-resonant, genuine $t\bar{t}j$ diagrams (see Figure \ref{fig:fig1}).
It is only in the narrow-width limit ($\Gamma_t/m_t \to 0$) that the additional contributions, part of the so-called \textit{off-shell effects}, are fully suppressed and let the cross section factorize into on-shell $t\bar{t}j$ production and decay. 

Despite technical advances, calculations involving multi-particle final states remain challenging. All previous studies of $t\bar{t}j$ production at the next-to-leading order (NLO) have resorted to the approximation of on-shell top quarks. This allowed significant progress in the state-of-the-art description, while being adequate for many applications. There are however issues that cannot be addressed without a complete calculation, like the impact of the non-resonant irreducible backgrounds and the relevance of the off-shellness of top quark and gauge boson decays. If, on the one hand, such effects are expected to be suppressed by powers of $\Gamma_t/m_t$ for inclusive observables \cite{Denner:2010jp,Bevilacqua:2010qb,Cascioli:2013wga,Denner:2012yc,Heinrich:2013qaa}, they are known to play a more relevant role in specific regions of the phase space.

The first QCD corrections to $t\bar{t}+\mbox{jet}$ hadroproduction have been computed in the picture of stable top quarks \cite{Dittmaier:2007wz,Dittmaier:2008uj}. Afterwards, effects of top quark decays have been included, first at LO \cite{Melnikov:2010iu} and then at NLO accuracy \cite{Melnikov:2011qx}. On-shell $t\bar{t}j$ production has been also matched with parton showers at NLO \cite{Kardos:2011qa, Alioli:2011as, Czakon:2015cla}. It is only quite recently that the QCD corrections to the full process, $pp \to e^+ \nu_e \mu^- \bar{\nu}_\mu b \bar{b} j + X$, have started to appear. We report on the first calculation of this kind, as presented in \cite{Bevilacqua:2015qha}.

\begin{figure}
\begin{center}
\begin{subfigure}{.23\textwidth}
  \centering
  \includegraphics[width=0.8\textwidth]{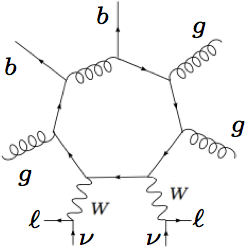}
  \caption{}
  \label{fig:sfig21}
\end{subfigure}
\begin{subfigure}{.23\textwidth}
  \centering
  \includegraphics[width=0.9\textwidth]{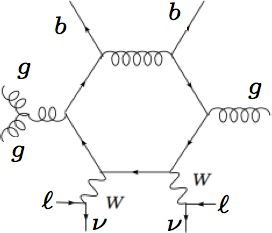}
  \caption{}
  \label{fig:sfig22}
\end{subfigure}
\begin{subfigure}{.23\textwidth}
  \centering
  \includegraphics[width=0.98\textwidth]{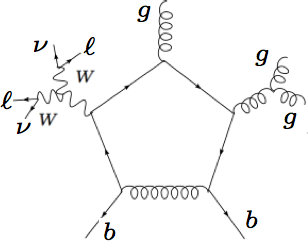}
  \caption{}
  \label{fig:sfig23}
\end{subfigure}
\begin{subfigure}{.23\textwidth}
  \centering
  \includegraphics[width=0.8\textwidth]{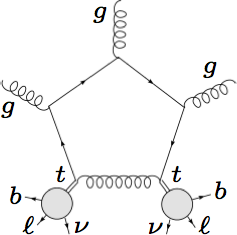}
  \caption{}
  \label{fig:sfig24}
\end{subfigure}
\caption{Representative one-loop contributions to $gg \to e^+ \nu_e \mu^- \bar{\nu}_\mu b \bar{b} g$ at the order $\mathcal{O}(\alpha_S^4 \alpha^4)$. Diagrams (a,b,c) are examples of non-factorizable contributions to $t\bar{t}+\mbox{jet}$ production.}
\label{fig:fig2}
\end{center}
\end{figure}

\section{Technical aspects of the calculation}

According to \texttt{QGRAF} \cite{Nogueira:1991ex} there are about 39000 one-loop diagrams contributing to the amplitude of the most challenging partonic subprocess, $gg \to e^+ \nu_e \mu^- \bar{\nu}_\mu b \bar{b} g$, at the order $\mathcal{O}(\alpha_S^4 \alpha^4)$. The most complicated ones are the 120 heptagons and 1155 hexagons, with tensor integrals up to rank six (see Figure \ref{fig:fig2}). We report these numbers as they customarily measure the complexity of NLO calculations, albeit we do not evaluate individual Feynman diagrams in our approach but rather employ more efficient Dyson-Schwinger recursion in association with the OPP reduction method \cite{Ossola:2006us,Ossola:2008xq,Draggiotis:2009yb}. The virtual corrections are computed by use of the packages  \texttt{HELAC-1LOOP} \cite{vanHameren:2009dr}, \texttt{CutTools} \cite{Ossola:2007ax} and \texttt{OneLOop} \cite{vanHameren:2010cp}, which are part of the \texttt{HELAC-NLO} framework \cite{Bevilacqua:2011xh}. Inheriting the structures of the \texttt{HELAC-PHEGAS} Monte Carlo \cite{Kanaki:2000ey,Papadopoulos:2000tt,Cafarella:2007pc}, the framework provides all the elements required to compute NLO QCD corrections to arbitrary processes in the SM. New functionalities have been introduced to cope with the complexity of the current project, among which the optimization of the algorithms for selecting loop topologies and the automated selection of contributions of different perturbative orders in $\alpha_S$ and $\alpha$. Numerical stability is monitored by checking Ward identities at every phase space point, using higher precision to recompute  events which fail the gauge-invariance check. To regularize resonances of unstable particles, the complex mass scheme \cite{Denner:2006ic} is employed. This requires the evaluation of scalar integrals with complex masses, which is  supported by the \texttt{OneLOop} library. The infrared divergencies arising from the real corrections are isolated by use of subtraction methods. Specifically, we adopt two independent schemes in our calculation: the standard Catani-Seymour subtraction \cite{Catani:1996vz,Catani:2002hc} and the alternative Nagy-Soper scheme \cite{Bevilacqua:2013iha}, both implemented in \texttt{HELAC-DIPOLES} \cite{Czakon:2009ss}. Phase space integrations are performed with the multichannel generator \texttt{KALEU} \cite{vanHameren:2010gg}. We cross check the stability of the real corrections by systematic comparisons of the results obtained with the two subtraction schemes.

\section{Phenomenological results}

We present here selected results of interest for the LHC Run I at the energy of 8 TeV. The SM parameters are set as follows,
\begin{eqnarray}
&& G_F = 1.16637 \cdot 10^{-5} \mbox{ GeV}, \;\;\; \; m_t = 173.3 \mbox{ GeV}, \nonumber \\
&& m_W = 80.399 \mbox{ GeV}, \;\;\; \;\;\; \;\;\; \;\;\; \;\;\; \;\; \Gamma_W = 2.09974 \mbox{ GeV}, \nonumber \\
&& m_Z = 91.1876 \mbox{ GeV}, \;\;\; \;\;\; \;\;\; \;\;\; \;\;\; \; \Gamma_Z = 2.50966 \mbox{ GeV}, \nonumber \\
&& \Gamma_t^{\mbox{\tiny{LO}}} = 1.48132 \mbox{ GeV}, \;\;\; \;\;\; \;\;\; \;\;\; \;\;\, \Gamma_t^{\mbox{\tiny{NLO}}} = 1.3542 \mbox{ GeV}. \nonumber
\end{eqnarray}
We consistently evaluate the top quark width at LO and NLO \cite{Jezabek:1988iv}. Since leptonic decays do not receive QCD corrections, the widths of $W$ and $Z$ bosons are the same everywhere in our calculation.
All leptons and quarks,  except the top, are considered massless.  We adopt the MSTW2008 parton distribution functions \cite{Martin:2009iq}, specifically MSTW2008lo68cl with 1-loop running $\alpha_s$ at LO and MSTW2008nlo68cl with 2-loop running $\alpha_s$ at NLO. Due to their small size (0.8\% of the total LO cross section), contributions from initial-state $b$ quarks are neglected. Jets are defined out of partons with pseudorapidity $\vert \eta \vert < 5$ using the \textit{anti}-$k_T$ clustering algorithm \cite{Cacciari:2008gp} with resolution parameter $R=0.5$. Our analysis requires exactly two $b$-jets, at least one light-jet, two charged leptons and missing $p_T$. The following phase space cuts are applied,
\begin{eqnarray}
&& p_{T_\ell}  > 30 \mbox{ GeV}, \;\;\;\; p_{T_j}  > 40 \mbox{ GeV}, \;\;\;\;  p_{T_{miss}}  > 40 \mbox{ GeV}, \nonumber \\
&& \Delta R_{jj}  > 0.5, \;\;\;\; \;\;\;\; \; \Delta R_{\ell\ell}  > 0.4, \;\;\;\; \;\;\;\; \;\, \Delta R_{\ell j}  > 0.4, \nonumber \\
&& \vert y_\ell \vert < 2.5, \;\;\;\; \;\;\;\; \;\;\;\; \; \vert y_j \vert < 2.5, \nonumber
\end{eqnarray}
where $\ell$ denotes charged leptons while $j$ stands for either light-jets or $b$-jets. For the renormalization and factorization scales we choose $\mu_R = \mu_F = \mu_0 = m_t$ and estimate scale uncertainties by a factor-2 variation around the central value $\mu_0$. 

\begin{figure}
\begin{center}
\includegraphics[width=.45\textwidth]{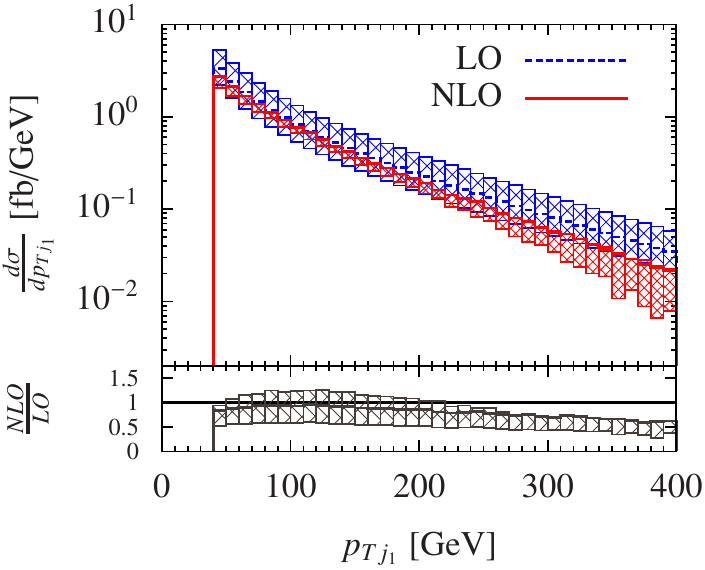} 
\hspace{0.01\textwidth}
\includegraphics[width=.45\textwidth]{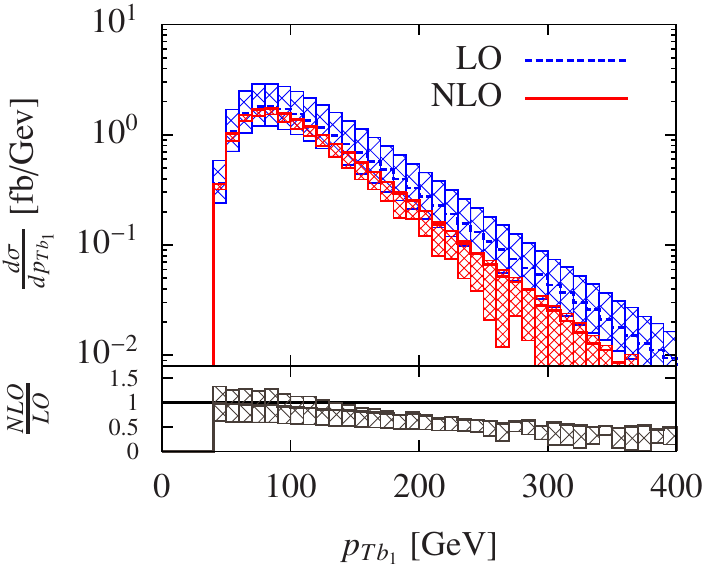} \\
\includegraphics[width=.431\textwidth]{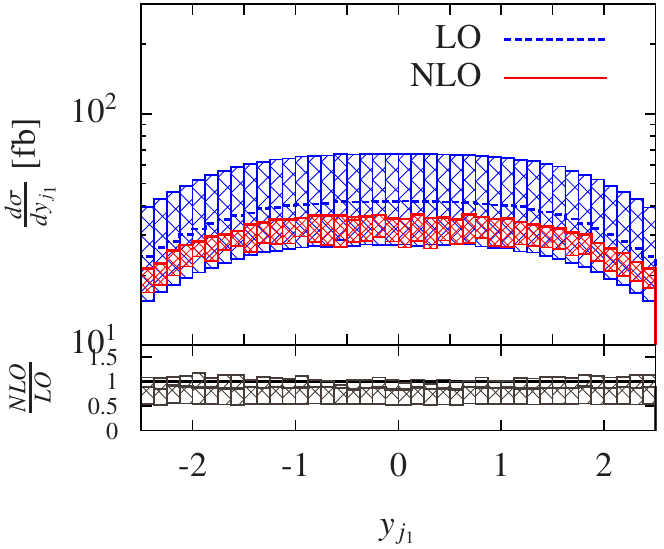} 
\hspace{0.026\textwidth}
\includegraphics[width=.431\textwidth]{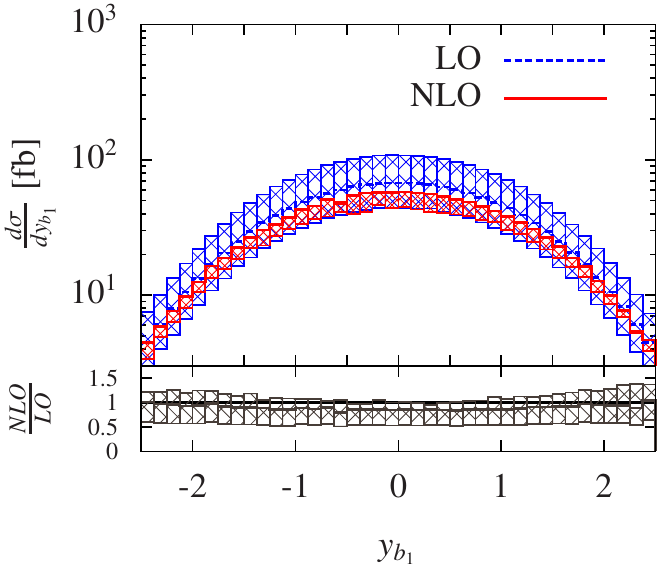}
\caption{\textit{Upper plots}: transverse momentum of the hardest light-jet (left) and of $b$-jet (right). \textit{Lower plots}; rapidity of the hardest light-jet (left) and $b$-jet (right). Results for $pp \to e^+ \nu_e \mu^- \bar{\nu}_\mu b \bar{b} j + X$ at the LHC with $\sqrt{s} = 8$ TeV. The bands denote estimates of scale uncertainties.}
\label{fig:figpTy}
\end{center}
\end{figure}

For the case of the LHC with $\sqrt{s} = 8$ TeV, our calculation gives the following results,
\begin{equation}
\sigma_{\mbox{\tiny{LO}}}  = 183.1^{\,+112.2}_{-64.2} \mbox{ fb} ,  \hspace{0.1\textwidth} \sigma_{\mbox{\tiny{NLO}}}  = 159.7 ^{\,-33.1}_{-7.9} \mbox{ fb} , 
\end{equation}
where the error bands denote scale uncertainties. We observe moderate, negative NLO QCD corrections of -13\% at the central scale choice $\mu_0 = m_t$. Also, the scale uncertainty of the total cross section is significantly reduced going from LO to NLO, from about 60\% down to 20\%. It is interesting to note that the higher-order corrections have a different impact on different observables. Figure \ref{fig:figpTy} shows distributions of transverse momentum and rapidity for the hardest light-jet and $b$-jet respectively. The upper panels contain the distributions themselves with the associated scale-dependence bands, the lower panels display the differential $K$ factor. While corrections look relatively stable for the case of  rapidities, shape distortions up to 50\% affect the $p_T$ distributions. Clearly, rescaling LO differential cross sections with a suitably chosen global $K$ factor is not a fair approximation of the full NLO result in this case. Further insight into judicious dynamical scales which could help to obtain more stable differential $K$ factors is required.

\begin{figure}
\begin{center}
\includegraphics[width=.45\textwidth]{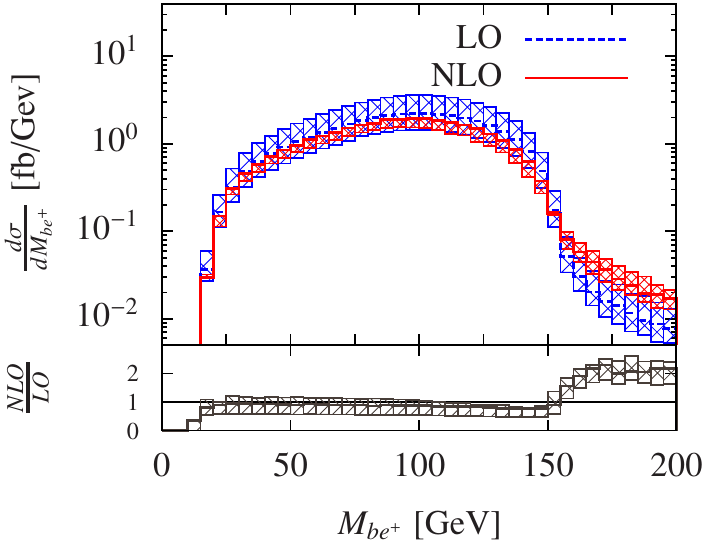}
\hspace{0.01\textwidth}
\includegraphics[width=.45\textwidth]{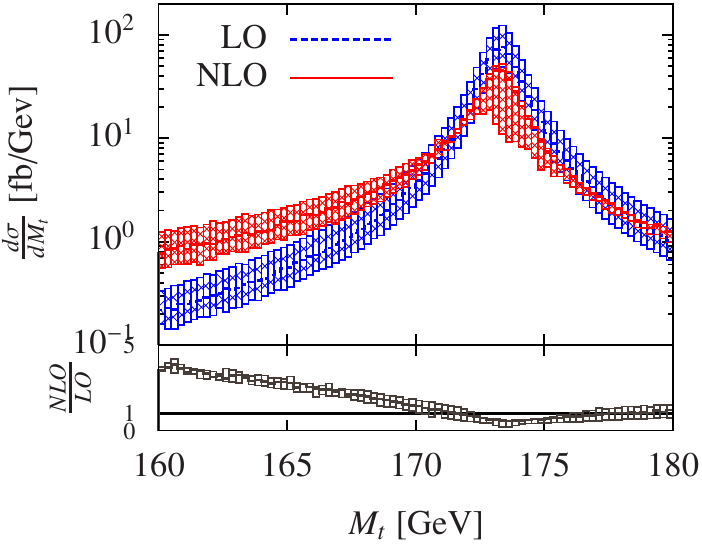} 
\caption{\textit{Left plot}: minimal $M_{be^+}$ invariant mass. \textit{Right plot}: invariant mass of the top quark reconstructed from its decay products. Results for $pp \to e^+ \nu_e \mu^- \bar{\nu}_\mu b \bar{b} j + X$ at the LHC with $\sqrt{s}=8$ TeV. The bands denote estimates of scale uncertainties.}
\label{fig:figM}
\end{center}
\end{figure}

To get an estimate of the numerical relevance of the non-factorizable corrections, we have also compared the results of the full calculation with its narrow-width limit. The latter is obtained by rescaling consistently the $tbW$ coupling and the top quark width by a small factor in order to mimic the limit $\Gamma_t \to 0$. Based on this procedure, we estimate the impact of the off-shell effects at the level of $1\% \, (2\%)$ of the total LO (NLO) cross section, fairly consistent with the value of the ratio $\Gamma_t/m_t$ which characterizes  the expected order of magnitude of such contributions at the inclusive level. It should be noticed, however, that the impact of the off-shell effects can be much larger on a more exclusive ground. Previous studies on $t\bar{t}$ production have shown that such effects reach several tens of percent in observables such as the cross section in exclusive $b$-jet bins \cite{Cascioli:2013wga}, or the minimum invariant mass of the positron and $b$-jet (hereafter denoted $M_{be^+}$) \cite{Denner:2012yc,Heinrich:2013qaa}. The latter is of particular interest, related as it is to one of the currently used methods for extracting the top quark mass \cite{Chatrchyan:2013boa,Aad:2015nba,Aaboud:2016igd}. The $M_{be^+}$ distribution for our $t\bar{t}j$ process is shown in Figure \ref{fig:figM}, together with the invariant mass of the top quark reconstructed from its decay products. Once more, the higher-order corrections are important to describe properly the whole range of these observables. The $M_{be^+}$ distribution displays the signature of a kinematical endpoint around the value $M_{be^+} = \sqrt{m_t^2-m_W^2} \approx 153.5$ GeV. This endpoint behaviour is ascribed to contributions from on-shell decays of top quarks and $W$ bosons. Additional jet radiation and off-shell effects smear this endpoint and generate a tail at large values of invariant mass, which is highly sensitive to QCD corrections.


\section{Conclusions}

We have computed NLO QCD corrections to the process $pp \to e^+ \nu_e \mu^- \bar{\nu}_\mu b \bar{b} j + X$, including for the first time complete off-shell effects for top and $W$ boson decays.
The QCD corrections look globally moderate (-13\% of the total cross section) but display a larger impact at the differential level. A thorough investigation of dynamical scales is desirable in order to improve the convergence of the perturbative expansion in several distributions of interest. We have estimated the size of the top quark off-shell effects at the level of 2\% of the total NLO cross section, in fair agreement with expectations dictated by the ratio $\Gamma_t/m_t$.
The results presented in this work are the starting point of a wider analysis aimed at providing more accurate NLO predictions for $t\bar{t}+\mbox{jet}$ production in the leptonic decay channel, without resorting to any on-shell approximation. Our results can help to improve the description of the $t\bar{t}+\mbox{jet}$ SM background for analyses of Higgs production via Vector Boson Fusion, $t\bar{t}H$ production or searches of signals beyond the SM. They have also applications to alternative methods for the  determination of the top quark mass \cite{Alioli:2013mxa,Aad:2015waa}, where a full simulation of the $t\bar{t}j$ final state beyond the narrow-width approximation is desirable for a more precise assessment of the theoretical uncertainties.

\end{document}